\documentclass{aa}
\usepackage{txfonts}  
\usepackage{natbib}
\bibpunct{(}{)}{;}{a}{}{,} 
\usepackage{graphicx}
\begin{document}


\title{Probable detection of H$_2$D$^+$ in the starless core
Barnard~68}

\author{M.~R. Hogerheijde\inst{1} 
  \and P. Caselli\inst{2,3}
  \and M. Emprechtinger\inst{4}
  \and F.~F.~S. van der Tak\inst{5,6}
  \and J. Alves\inst{7,8}
  \and A. Belloche\inst{6}
  \and R. G\"usten\inst{6}
  \and A.~A. Lundgren\inst{9}
  \and L.-{\AA}. Nyman\inst{9}
  \and N. Volgenau\inst{4}
  \and M.~C. Wiedner\inst{4}
}

\offprints{M. Hogerheijde, \email{michiel@strw.leidenuniv.nl}}

\institute{Leiden Observatory, P.O. Box 9513, 2300 RA, Leiden, The
  Netherlands 
  \and INAF-Osservatorio Astrofisico di Arcetri, Largo E. Fermi 5,
  50125 Firenze, Italy
  \and Harvard-Smithsonian Center for Astrophysics, 60 Garden Street,
  Cambridge, MA 02138, USA
  \and I. Physikalisches Institut, Universit\"at zu K\"oln,
  Z\"ulpicher Stra{\ss}e 77, 50937 K\"oln, Germany
  \and Netherlands Institute for Space Research (SRON), P.O. Box 800,
  9700 AV, Groningen, The Netherlands
  \and Max-Planck-Institut f\"ur Radioastromie, Auf dem H\"ugel 69,
  53121 Bonn, Germany
  \and European Southern Observatory, Karl Schwarzschild-Stra{\ss}e 2,
  85748 Garching bei M\"unchen, Germany
  \and Present address: Centro Astron\'omico Hispano Alem\'an, Apt.\ 511, 04080
  Almer\'{\i}a, Spain
  \and European Southern Observatory, Casilla 19001, Santiago 19, Chile}

\date{Received / Accepted }

\abstract
{ The presence of H$_2$D$^+$ in dense cloud cores underlies
  ion-molecule reactions that strongly enhance the deuterium
  fractionation of many molecular species.}
{We determine the H$_2$D$^+$ abundance in one starless core,
  Barnard~68, that has a particularly well established physical,
  chemical, and dynamical structure. }
{We observed the ortho-H$_2$D$^+$ ground-state line
  $1_{10}$--$1_{11}$, the N$_2$H$^+$ $J$=4--3 line, and the
  H$^{13}$CO$^+$ 4--3 line with the APEX telescope.}
{We report the probable detection of the o-H$_2$D$^+$ line at
  an intensity $T_{\rm mb}$=$0.22\pm 0.08$~K and exclusively thermal
  line width, and find only upper limits to the N$_2$H$^+$ 4--3 and
  H$^{13}$CO$^+$ 4--3 intensities.}
{Within the uncertainties in the chemical reaction rates and the
  collisional excitation rates, chemical model calculations and
  excitation simulations reproduce the observed intensities and that
  of o-H$_2$D$^+$ in particular.}

\keywords{ISM: abundances -- ISM: individual objects: Barnard~68 --
  ISM: molecules -- Submillimeter}

\titlerunning{Probable detection of H$_2$D$^+$ in Barnard~68}
\authorrunning{Hogerheijde et al.}
\maketitle

\section{Introduction\label{s:intro}}

The densest and coldest cores of interstellar molecular clouds are
receiving much attention as the possible precursors of star
formation. Often referred to as starless or pre-protostellar cores,
they are characterized chemically by a large depletion of molecules as
they freeze out onto dust grains, and an associated increase in the
relative abundance of deuterated isotopomers of numerous species
\citep[e.g.,][]{kuiper:l1498, ceccarelli1998, loinard2002,
  bacmann2003, caselli:l1544_h2d+, stark2004}.  This increase is
explained by the reaction ${\rm H_3^+ + HD \getsto H_2D^+ + H_2}$,
which is favored in the forward direction at low temperatures, and
subsequent ion-molecule reactions involving H$_2$D$^+$. Detection of
H$_2$D$^+$ in dense cloud cores directly tests this proposed
mechanism. Through its spectral line shape it also probes the core's
velocity field, in regions where all other molecular tracers are
strongly depleted. First predicted by \citet{dalgarno:h2d+}, the
ground-state transition of ortho-H$_2$D$^+$, $1_{10}$--$1_{11}$, lies
in the submillimeter wavelength range near 372~GHz, in a region with
poor atmospheric transmission. Under excellent observing conditions on
Mauna Kea (Hawai`i), H$_2$D$^+$ has been successfully detected toward
astronomical sources, including the young stellar object
\object{NGC~1333~IRAS~4A} \citep{stark:n1333_h2d+}, the starless core
\object{L1544} \citep{caselli:l1544_h2d+}, and the circumstellar disks
of \object{TW~Hya} and \object{DM~Tau}
(\citealt{ceccarelli:twhya_dmtau_h2d+}; but also see
\citealt{guilloteau:dmtau}) . This Letter reports the probable
detection of the H$_2$D$^+$ $1_{10}$--$1_{11}$ line toward the
well-studied starless core \object{Barnard~68} conducted with the
Atacama Pathfinder EXperiment (APEX\footnote{This publication is based
  on data acquired with the Atacama Pathfinder EXperiment (APEX). APEX
  is a collaboration of the Max-Planck-Institut f\"ur Radioastronomie,
  the European Southern Observatory, and the Onsala Space
  Observatory.})  telescope on Chajnantor (Chile).

Barnard~68 (B68) is one of the most extensively studied
starless cores. Using stellar extinction measurements, its density
structure has been found to be well matched by a near-critical
Bonnor-Ebert sphere \citep{alves:b68}. Bonnor-Ebert spheres
\citep{ebert,bonnor} describe the equilibrium configuration of
self-gravitating cloud cores just before the onset of
collapse. \citet{bergin:b68} found strong depletion of C$^{18}$O in
B68.
\citet{lada:b68} analyzed the velocity field of B68, and found that
the line widths in the center are close to thermal, leaving no room for
significant turbulent motion. In the outskirts of the core, line
centroid shifts suggest a non-radial pulsating motion.

With the successful commissioning of APEX, the 372~GHz ground-state
transition of ortho-H$_2$D$^+$ has come within reach of regular
observing, given the good local weather conditions. This Letter
presents the probable detection of H$_2$D$^+$ toward the center of
this starless core obtained during the Science Verification of APEX
(\S \ref{s:obs}) and discusses the emission strength in the framework
of a chemical model including depletion and deuteration (\S
\ref{s:discussion}).

\section{Observations and results\label{s:obs}}

The APEX telescope observed the H$_2$D$^+$ $1_{10}$--$1_{11}$ line at
372.421385~GHz on 2005 July 24 and 25 using the APEX-2a receiver and
the FFTS backend with a bandwidth of 1~GHz and 16384 channels. This
frequency setting also covers the N$_2$H$^+$ $J$=4--3 line at
372.6725090~GHz. The telescope was pointed at the $A_V$ peak measured
by \cite{alves:b68} at $\alpha_{2000}=17^{\rm h}22{\rm m}38{\fs}6$ and
$\delta_{2000}=-23\degr 49' 46{\farcs}0$. The observations
  were taken in position-switched mode, using an emission-free
  reference position. Pointing was checked on the nearby object
RAFGL~1922. During the observations the source was at elevations of
$25^\circ$--$40^\circ$. A precipitable water vapor column of 0.47~mm
resulted in DSB system temperatures of 150--250~K. On 2005 July 25,
the H$^{13}$CO$^+$ $J$=4--3 line at 346.998546~GHz was observed with a
similar set up. After the observations, the velocity scale of the
spectra was recalculated by the telescope staff to correct for a small
(0.1~km~s$^{-1}$) inaccuracy during data taking.

After careful inspection of the data, the individual 30~sec scans were
averaged using the CLASS software package; a total integration time
(on+off) of 36~min was obtained for the H$_2$D$^+$ line (12.6~min for
H$^{13}$CO$^+$). A sinusoidal baseline with a period of $\sim
350$~km~s$^{-1}$ was removed, followed by a first order baseline in
the area surrounding the expected line. After smoothing the spectral
resolution to 0.098~km~s$^{-1}$, a rms noise level of 0.059~K on the
$T_A^*$ scale was found (0.058~K for H$^{13}$CO$^+$ in
0.11~km~s$^{-1}$ channels), adopting a forward efficiency of 0.97. The
data were subsequently transformed to the main beam antenna
temperature scale by division by 0.73, the recommended mean beam
efficiency in the 345~GHz band \citep{guesten:apex}.

\begin{figure}
 \resizebox{\hsize}{!}{\includegraphics{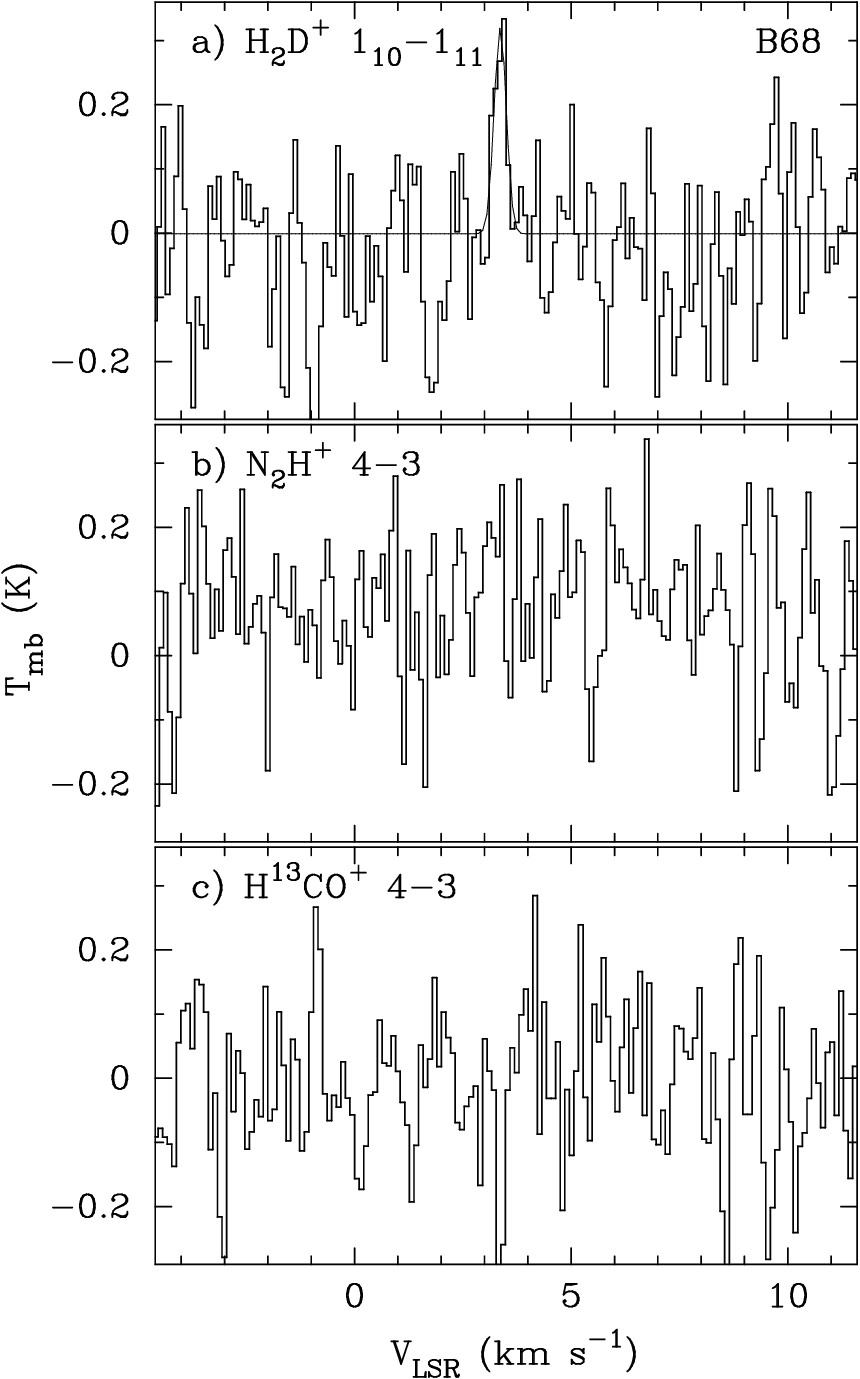}}
 \caption{{\bf (a)} Detection spectrum of H$_2$D$^+$
  $1_{10}$--$1_{11}$ toward the center of B68 (histogram), with the
  best-fit Gaussian line profile superposed (thin line). The model of
  \S\ref{s:discussion} yields an identical profile for the modified
  collisional excitation rates as discussed. {\bf (b)} Non-detection
  of N$_2$H$^+$ 4--3 at the same position. {\bf (c)} Non-detection of
  H$^{13}$CO$^+$ 4--3 at the same position.}
 \label{f:spec}
\end{figure}

\begin{figure*}
 \resizebox{\hsize}{!}{\includegraphics{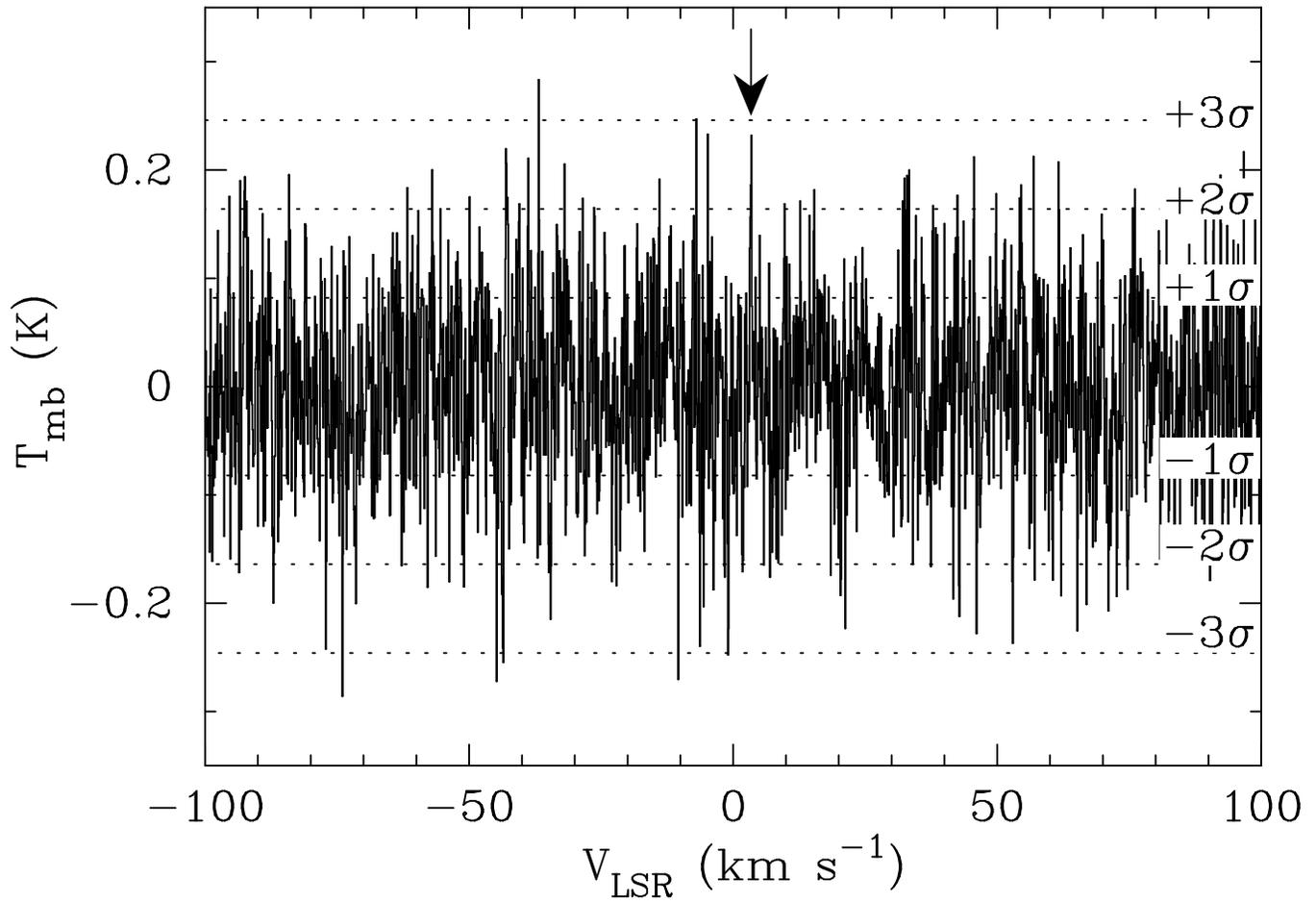}}
 \caption{Detection spectrum of H$_2$D$^+$ $1_{10}$--$1_{11}$ toward
  the center of B68 over a baseline range of 200~km~s$^{-1}$. Noise
  levels at $\pm 1\sigma$, $\pm 2\sigma$, and $\pm 3\sigma$ are
  indicated. The arrow indicates the $V_{\rm LSR}$ of B68 of
  3.36~km~s$^{-1}$.}
 \label{f:fullspec}
\end{figure*}

The H$_2$D$^+$ line was detected at a signal-to-noise ratio of $\sim$3
(Fig.~\ref{f:spec}). Fitting a single Gaussian line to the spectrum
yields $T_{\rm mb}$=$0.222\pm 0.082$~K, $V_{\rm LSR}$=$3.36\pm
0.05$~km~s$^{-1}$, $\Delta V$=$0.33\pm 0.1$~km~s$^{-1}$, and an
integrated line intensity of $0.078\pm 0.015$~K~km~s$^{-1}$. Although
the line peak is detected at only $2.7\sigma$, the integrated
intensity is detected at $5.2\sigma$, and we argue that the detection
of H$_2$D$^+$ in B68 is probable (only deeper integration can make the
result more secure). Fig.~\ref{f:fullspec} in the on-line material
shows the H$_2$D$^+$ spectrum over a 200~km~s$^{-1}$ range. Out
of the 1018 channels in this part of the spectrum, only 4 exceed the
$3\sigma$ level (0.4\%), as statistically expected. The significance
of the detection is further supported by the close match to the source
$V_{\rm LSR}$ of 3.36--3.37~km~s$^{-1}$ and the purely thermal line
width of 0.33~km~s$^{-1}$ (H$_2$D$^+$ at 10~K), as found by
\cite{lada:b68} for N$_2$H$^+$ and C$^{18}$O. All other peaks are much
narrower (1--2 channels).
%
%
Because H$_2$D$^+$ retains a high abundance at the center of the core
(see below), the apparent absence of turbulent motion in the
H$_2$D$^+$ line provides strong support for the conclusion of
\cite{lada:b68} that B68 is exclusively thermally supported. This
situation is very different from the velocity field in, e.g., L1544,
which shows significant velocity gradients and wider H$_2$D$^+$ lines
\citep{vdtak:h2d+}. No detection was made of either the N$_2$H$^+$
4--3 line to a 2$\sigma$ upper limit of 0.16~K or the H$^{13}$CO$^+$
4--3 line to 2$\sigma$ of 0.18~K. An emission peak at the correct
$V_{\rm LSR}$ for the N$_2$H$^+$ transition and the expected thermal
width of 0.075~km~s$^{-1}$ is likely noise.

In the documents accompanying the data release, it was noted that
lines tuned in the upper sideband of APEX-2a could be too strong by
40\%. This effect has been shown to be present for $^{12}$CO $J$=3--2,
and is thought to be due to a resonance near the CO line frequency. It
is therefore unlikely that our H$_2$D$^+$ observations suffer from
this effect, although no sideband-ratio measurements are available to
confirm this.

\section{Discussion\label{s:discussion}}

At the detected line strength of $0.22\pm 0.08$~K, the H$_2$D$^+$ line
is weaker than detections toward other starless cores such as L1544
($\sim 1$~K). Two aspects distinguish B68 from L1544. First, B68's
H$_2$ column density is lower than that of L1544, at $3.6\times
10^{22}$~cm$^{-2}$ and $(6$--$13)\times 10^{22}$~cm$^{-2}$
\citep{alves:b68,wardthompson:isophot,wardthompson:scuba}. Secondly,
B68's central density of $2\times 10^5$~cm$^{-3}$ is lower by a factor
of 6 than that of L1544. Given that the critical density of the
ground-state transition of ortho-H$_2$D$^+$ is $\sim 2\times
10^6$~cm$^{-3}$ (for our adopted collision rate; see below), a careful
analysis is required to test whether the decreased line strength is
due to (a combination of) sub-thermal excitation, lower total column
density, or lower abundance of this key deuterated species.

From the strength of the H$_2$D$^+$ line, assuming optically thin
conditions, we derive a beam-averaged column density of
$N$(H$_2$D$^+$)=$(1.5\pm 0.14)\times 10^{12}$~cm$^{-2}$ for an
excitation temperature of 10~K. This corresponds to thermal
equilibrium excitation at the kinetic temperature in B68. Given the
critical density quoted above, it is highly likely that the excitation
is subthermal, requiring a larger column density to reproduce the
emission. For example, an excitation temperature of 5~K implies
$N$(H$_2$D$^+$)=$(1.0\pm 0.3)\times 10^{13}$~cm$^{-2}$. This column
density corresponds to an average H$_2$D$^+$ abundance with respect to
H$_2$ of
$(2.9\pm 0.9)\times 10^{-10}$.

Further insight in the H$_2$D$^+$ abundance can be obtained
  from modeling the chemistry in the B68 core followed by a
  statistical equilibrium calculation of the H$_2$D$^+$ excitation and
  line radiative transfer. Following \citet{vastel} we use the simple
  chemical model of \citet{caselli:l1544_2}, updated to include the
  multiply deuterated forms of H$_3^+$ and the new values of the
  binding energies of CO and N$_2$ \citep{oberg:n2}. B68 is modeled
  as a spherical cloud with radius 12,000~AU and the density profile
  fitted to the Bonnor-Ebert solution by
  \citet{tafalla2002,tafalla2004}. The temperature profile follows
  that of L1544 \citep{young2004}, resulting in a temperature gradient
  from 13~K at the core edge to 9.8~K at its center.


The chemical model starts with fully undepleted abundances of
CO, N$_2$, and O, and lets the species interact with and freeze-out on
the dust grains which follow a MRN size distribution \citep{mrn} with a
minimum grain size of 0.005~$\mu$m. Assuming that the ion chemistry is
fast compared to depletion, the molecular ions follow from the
instantaneous neutral abundances. Following \citet{vastel} a
cosmic-ray ionization rate of $1.3\times 10^{-17}$~s$^{-1}$ and
sticking coefficient to $S=1.0$ \citep{burke1983} are
adopted. Finally, for H$_2$D$^+$ an ortho-to-para ratio of 1.0 is
adopted as expected at low temperatures \citep{gerlich2002}. We note
that the results of these calculations may depend quite sensitively on
the chemical reaction rates and their dependence on temperature
(Emprechtinger et al. in preparation), including the H$_2$D$^+$
destruction reaction with H$_2$ (\citealt{schlemmer2005} and Asvany et
al., in preparation).

\begin{figure}
 \resizebox{\hsize}{!}{\includegraphics{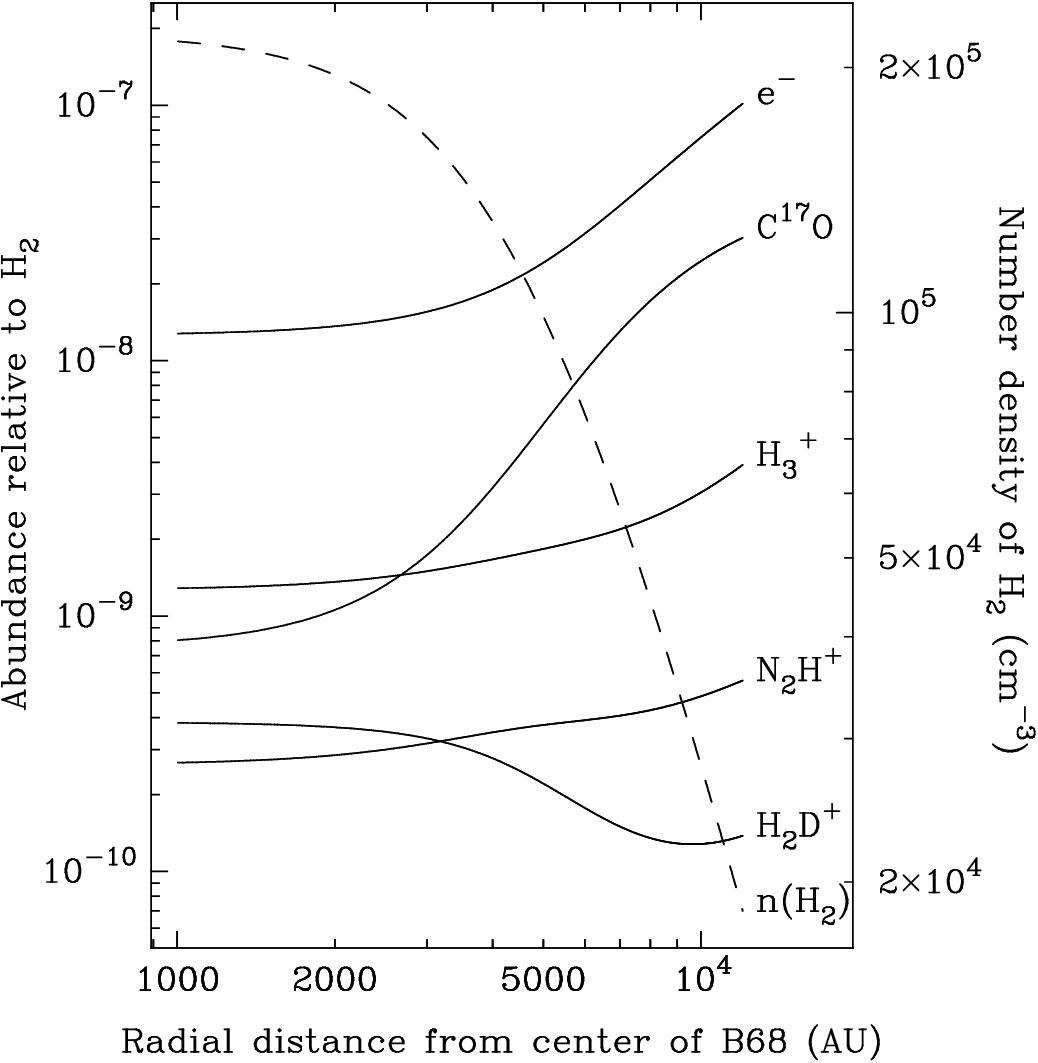}}
 \caption{Predicted abundances of several key species in the
 H$_2$D$^+$ modeling (solid lines) and H$_2$ number density (dashed
 line) as functions of radius in B68.}
 \label{f:mdl}
\end{figure}

Figure~\ref{f:mdl} shows the resulting abundance profiles of some
relevant species. Typical H$_2$D$^+$ abundances with respect to H$_2$
at the center of B68 are predicted to be $4\times 10^{-10}$. The
corresponding H$_2$D$^+$ column density is $5.0\times
10^{12}$~cm$^{-2}$, close to the beam-averaged column density
of $1\times 10^{13}$~cm$^{-2}$ found from the simple analysis above.

With these abundance profiles, the statistical equilibrium excitation
is calculated using the code of \cite{hogerheijde:amc}. The velocity
field is assumed to be static with only thermal broadening and no
turbulent motion. We ignore the significant motion in the outer
regions of the core reported by \citet{lada:b68}, since our
observations are taken to the core's center and the observed spectrum
shows no indication of line broadening.  The resulting spectrum is
convolved in a $17''$ FWHM beam appropriate for APEX at these
wavelengths, yielding intensities on the $T_{\rm mb}$ scale; a
distance of 100~pc is adopted for B68.


These calculations require reliable collisional rate coefficients,
which are not available in the literature. We follow \cite{vdtak:h2d+}
in adopting estimated rates, and investigate the effect of their
inherent uncertainty. Using these rates, we find a H$_2$D$^+$
$1_{10}$--$1_{11}$ intensity of 0.10~K, lower by a factor of 2 than
observed. Increasing the collisional rates by a factor of three,
entirely within the estimated uncertainty, increases the emergent
intensity to 0.24~K.  Varying the rates by factors of 10 up or down
produces a range of intensities of 0.011--0.48~K. The opacity of the
line is 0.13 and the excitation temperature ranges from 3 to 6~K,
consistent with the simple assumptions made above.


We conclude that the observed intensity of H$_2$D$^+$ is consistent
with chemical model predictions, but that the uncertainty in the
available collisional rate coefficients precludes a detailed
comparison, especially for subthermal excitation. 
Determination of more reliable rates is clearly
warranted. \citet{vdtak:h2d+} reach the same conclusion, including a
preference for collision rates enhanced by factors 3--10; like
these authors, we refrain from recommending adjusting the collision
rates because of the uncertainties in the chemical modeling.

Using collisional rates from the Leiden data base \citep{lambda}, our
model calculations are compatible with the upper limits on
H$^{13}$CO$^+$ and N$_2$H$^+$ 4--3 reported here, as well the
N$_2$H$^+$ and C$^{18}$O 1--0 detections of \cite{bergin:b68}
and N$_2$H$^+$ 3--2 from \cite{crapsi}. Quantitatively, we
find 19~mK for H$^{13}$CO$^+$ 4--3 (observed: $<0.15$~K, $2\sigma$);
0.79~K~km~s$^{-1}$ for C$^{18}$O 1--0 (0.85~K~km~s$^{-1}$). Our model
overproduces the N$_2$H$^+$ emission by about a factor of 2; we find
0.38~K for N$_2$H$^+$ 4--3 (observed $<0.16$~K, $2\sigma$), and values
of 3.8~K~km~s$^{-1}$ for N$_2$H$^+$ 1--0 (observed, 2.5~K~km~s$^{-1}$)
and 0.35~K~km~s$^{-1}$ for 3--2 (observed,
0.17~K~km~s$^{-1}$).


We conclude that our detection of H$_2$D$^+$ toward B68 is consistent
with models for the deuterium chemistry in starless cores which also
reproduces other observed lines. However, the lack of reliable
collision rates precludes any stronger statements about the H$_2$D$^+$
chemistry. Given the pivotal role of H$_2$D$^+$ in deuterium chemistry
and the well studied nature of B68, calculation of such rates is
warranted, especially now that observations of H$_2$D$^+$ ground-state
transition are possible with APEX. In the future, ALMA observations
of the resolved emission of H$_2$D$^+$ and other species will provide
further insight into the chemical state of B68, and only with reliable
collision rates will quantitative analysis of these observations be
possible.

\begin{acknowledgements}
The authors are indebted to the staff of the APEX telescope for their
efforts before, during, and after the observations. PC would like to
thank C.\ Ceccarelli for her fundamental contribution in the search for
H$_2$D$^+$ in prestellar cores. PC acknowledges support from the MIUR
grant ``Dust particles as factor of galactic evolution''. MRH's
research is supported by a VIDI grant from the Nederlandse Organisatie
voor Wetenschappelijk Onderzoek. The referee, L.~Pagani, is thanked
for his constructive comments that improved the paper.
\end{acknowledgements}


\begin{thebibliography}{31}
\expandafter\ifx\csname natexlab\endcsname\relax\def\natexlab#1{#1}\fi

\bibitem[{{Alves} {et~al.}(2001){Alves}, {Lada}, \& {Lada}}]{alves:b68}
{Alves}, J.~F., {Lada}, C.~J., \& {Lada}, E.~A. 2001, \nat, 409, 159

\bibitem[{{Bacmann} {et~al.}(2003){Bacmann}, {Lefloch}, {Ceccarelli},
  {Steinacker}, {Castets}, \& {Loinard}}]{bacmann2003}
{Bacmann}, A., {Lefloch}, B., {Ceccarelli}, C., {et~al.} 2003, \apjl, 585, L55

\bibitem[{{Bergin} {et~al.}(2002){Bergin}, {Alves}, {Huard}, \&
  {Lada}}]{bergin:b68}
{Bergin}, E.~A., {Alves}, J., {Huard}, T., \& {Lada}, C.~J. 2002, \apjl, 570,
  L101

\bibitem[{{Bonnor}(1958)}]{bonnor}
{Bonnor}, W.~B. 1958, \mnras, 118, 523

\bibitem[{{Burke} \& {Hollenbach}(1983)}]{burke1983}
{Burke}, J.~R. \& {Hollenbach}, D.~J. 1983, \apj, 265, 223

\bibitem[{{Caselli} {et~al.}(2003){Caselli}, {van der Tak}, {Ceccarelli}, \&
  {Bacmann}}]{caselli:l1544_h2d+}
{Caselli}, P., {van der Tak}, F.~F.~S., {Ceccarelli}, C., \& {Bacmann}, A.
  2003, \aap, 403, L37

\bibitem[{{Caselli} {et~al.}(2002){Caselli}, {Walmsley}, {Zucconi}, {Tafalla},
  {Dore}, \& {Myers}}]{caselli:l1544_2}
{Caselli}, P., {Walmsley}, C.~M., {Zucconi}, A., {et~al.} 2002, \apj, 565, 344

\bibitem[{{Ceccarelli} {et~al.}(1998){Ceccarelli}, {Castets}, {Loinard},
  {Caux}, \& {Tielens}}]{ceccarelli1998}
{Ceccarelli}, C., {Castets}, A., {Loinard}, L., {Caux}, E., \& {Tielens},
  A.~G.~G.~M. 1998, \aap, 338, L43

\bibitem[{{Ceccarelli} {et~al.}(2004){Ceccarelli}, {Dominik}, {Lefloch},
  {Caselli}, \& {Caux}}]{ceccarelli:twhya_dmtau_h2d+}
{Ceccarelli}, C., {Dominik}, C., {Lefloch}, B., {Caselli}, P., \& {Caux}, E.
  2004, \apjl, 607, L51

\bibitem[{Crapsi} {et~al.}(2005)]{crapsi} {Crapsi}, A., {Caselli}, P.,
{Walmsley}, C.~M., {Myers}, P.~C., {Tafalla}, M., {Lee}, C.~W., \&
{Bourke}, T.~L.\ 2005, \apj, 619, 379

\bibitem[{{Dalgarno} {et~al.}(1973){Dalgarno}, {Herbst}, {Novick}, \&
  {Klemperer}}]{dalgarno:h2d+}
{Dalgarno}, A., {Herbst}, E., {Novick}, S., \& {Klemperer}, W. 1973, \apjl,
  183, L131

\bibitem[{{Ebert}(1957)}]{ebert}
{Ebert}, R. 1957, Zeitschrift f\"ur Astrophysik, 42, 263

\bibitem[{{Gerlich} {et~al.}(2002){Gerlich}, {Herbst}, \&
  {Roueff}}]{gerlich2002}
{Gerlich}, D., {Herbst}, E., \& {Roueff}, E. 2002, \planss, 50, 1275

\bibitem[{{Guilloteau} {et~al.}(2006){Guilloteau}, {Pi{\'e}tu}, {Dutrey}, \&
  {Gu{\'e}lin}}]{guilloteau:dmtau}
{Guilloteau}, S., {Pi{\'e}tu}, V., {Dutrey}, A., \& {Gu{\'e}lin}, M. 2006,
  \aap, 448, L5

\bibitem[{{G\"usten} {et~al.}(2006)}]{guesten:apex}
{G\"usten}, R. {et al.} 2006, A\&A, this volume

\bibitem[{{Hogerheijde} \& {van der Tak}(2000)}]{hogerheijde:amc}
{Hogerheijde}, M.~R. \& {van der Tak}, F.~F.~S. 2000, \aap, 362, 697

\bibitem[{{Kuiper} {et~al.}(1996){Kuiper}, {Langer}, \&
  {Velusamy}}]{kuiper:l1498}
{Kuiper}, T.~B.~H., {Langer}, W.~D., \& {Velusamy}, T. 1996, \apj, 468, 761

\bibitem[{{Lada} {et~al.}(2003){Lada}, {Bergin}, {Alves}, \&
  {Huard}}]{lada:b68}
{Lada}, C.~J., {Bergin}, E.~A., {Alves}, J.~F., \& {Huard}, T.~L. 2003, \apj,
  586, 286

\bibitem[{{Loinard} {et~al.}(2002){Loinard}, {Castets}, {Ceccarelli},
  {Lefloch}, {Benayoun}, {Caux}, {Vastel}, {Dartois}, \&
  {Tielens}}]{loinard2002}
{Loinard}, L., {Castets}, A., {Ceccarelli}, C., {et~al.} 2002, \planss, 50,
  1205

\bibitem[{{Mathis} {et~al.}(1977){Mathis}, {Rumpl}, \& {Nordsieck}}]{mrn}
{Mathis}, J.~S., {Rumpl}, W., \& {Nordsieck}, K.~H. 1977, \apj, 217, 425

\bibitem[{{{\"O}berg} {et~al.}(2005){{\"O}berg}, {van Broekhuizen}, {Fraser},
  {Bisschop}, {van Dishoeck}, \& {Schlemmer}}]{oberg:n2}
{{\"O}berg}, K.~I., {van Broekhuizen}, F., {Fraser}, H.~J., {et~al.} 2005,
  \apjl, 621, L33

\bibitem[{{Schlemmer} {et~al.}(2005){Schlemmer}, {Asvany}, \&
  {Hugo}}]{schlemmer2005}
{Schlemmer}, S., {Asvany}, O., \& {Hugo}, E. 2005, in Proc. IAU Symp. 231, eds.
  D. Lis, G. A. Blake, E. Herbst (Cambridge University Press), 223

\bibitem[{{Sch{\"o}ier} {et~al.}(2005){Sch{\"o}ier}, {van der Tak}, {van
  Dishoeck}, \& {Black}}]{lambda}
{Sch{\"o}ier}, F.~L., {van der Tak}, F.~F.~S., {van Dishoeck}, E.~F., \&
  {Black}, J.~H. 2005, \aap, 432, 369

\bibitem[{{Stark} {et~al.}(2004){Stark}, {Sandell}, {Beck}, {Hogerheijde}, {van
  Dishoeck}, {van der Wal}, {van der Tak}, {Sch{\"a}fer}, {Melnick}, {Ashby},
  \& {de Lange}}]{stark2004}
{Stark}, R., {Sandell}, G., {Beck}, S.~C., {et~al.} 2004, \apj, 608, 341

\bibitem[{{Stark} {et~al.}(1999){Stark}, {van der Tak}, \& {van
  Dishoeck}}]{stark:n1333_h2d+}
{Stark}, R., {van der Tak}, F.~F.~S., \& {van Dishoeck}, E.~F. 1999, \apjl,
  521, L67

\bibitem[{{Tafalla} {et~al.}(2004){Tafalla}, {Myers}, {Caselli}, \&
  {Walmsley}}]{tafalla2004}
{Tafalla}, M., {Myers}, P.~C., {Caselli}, P., \& {Walmsley}, C.~M. 2004, \aap,
  416, 191

\bibitem[{{Tafalla} {et~al.}(2002){Tafalla}, {Myers}, {Caselli}, {Walmsley}, \&
  {Comito}}]{tafalla2002}
{Tafalla}, M., {Myers}, P.~C., {Caselli}, P., {Walmsley}, C.~M., \& {Comito},
  C. 2002, \apj, 569, 815

\bibitem[{{van der Tak} {et~al.}(2005){van der Tak}, {Caselli}, \&
  {Ceccarelli}}]{vdtak:h2d+}
{van der Tak}, F.~F.~S., {Caselli}, P., \& {Ceccarelli}, C. 2005, \aap, 439,
  195

\bibitem[{{Vastel} {et~al.}(2006){Vastel}, {Caselli}, {Ceccarelli}, {Phillips},
  {Wiedner}, {Peng}, {Houde}, \& {Dominik}}]{vastel}
{Vastel}, C., {Caselli}, P., {Ceccarelli}, C., {et~al.} 2006, ApJ, in press

\bibitem[{{Ward-Thompson} {et~al.}(2002){Ward-Thompson}, {Andr{\'e}}, \&
  {Kirk}}]{wardthompson:isophot}
{Ward-Thompson}, D., {Andr{\'e}}, P., \& {Kirk}, J.~M. 2002, \mnras, 329, 257

\bibitem[{{Ward-Thompson} {et~al.}(1999){Ward-Thompson}, {Motte}, \&
  {Andr\'e}}]{wardthompson:scuba}
{Ward-Thompson}, D., {Motte}, F., \& {Andr\'e}, P. 1999, \mnras, 305, 143

\bibitem[{{Young} {et~al.}(2004){Young}, {Lee}, {Evans}, {Goldsmith}, \&
  {Doty}}]{young2004}
{Young}, K.~E., {Lee}, J.-E., {Evans}, N.~J., {Goldsmith}, P.~F., \& {Doty},
  S.~D. 2004, \apj, 614, 252

\end{thebibliography}

\Online

\end{document}